\documentclass[twocolumn]{aastex701}

\usepackage[flushleft]{threeparttable}
\usepackage{adjustbox}
\usepackage{amsmath}

\begin{document}

\title{Constraining the Inner Accretion Geometry of GX~9+9 with Broadband Spectral and Timing Studies}

\author[0000-0003-3499-9273]{Swadesh Chand}
\affiliation{Institute of Astronomy, National Tsing Hua University, Hsinchu 300044, Taiwan}
\email[show]{swadesh.chand@mx.nthu.edu.tw; swadesh.chand@gmail.com}

\author[0000-0002-6051-6928]{Srimanta Banerjee}
\affiliation{Istituto Nazionale di Astrofisica (INAF) – IASF Palermo, Via U. La Malfa 153, I-90146 Palermo, Italy}
\email{}

\author[0000-0002-5617-3117]{Hsiang-Kuang Chang}
\affiliation{Institute of Astronomy, National Tsing Hua University, Hsinchu 300044, Taiwan}
\affiliation{Department of Physics, National Tsing Hua University, Hsinchu 300044, Taiwan}
\affiliation{Institute of Space Engineering, National Tsing Hua University, Hsinchu 300044, Taiwan}
\affiliation{Center for Theory-Computation-Data Science Research (CTCD), National Tsing Hua University, Hsinchu 300044, Taiwan}
\email{}

\author[0000-0002-0333-2452]{Andrzej A. Zdziarski}
\affiliation{Nicolaus Copernicus Astronomical Center, Polish Academy of Sciences, Bartycka 18, PL-00-716 Warszawa, Poland}
\email{}

\author[0000-0003-1589-2075]{Gulab C. Dewangan}
\affiliation{Inter-University Centre for Astronomy and Astrophysics, Pune, Maharashtra-411007, India}
\email{}

\author{Vivek K. Agrawal}
\affiliation{Space Astronomy Group, ISITE Campus, U. R. Rao Satellite Center, Outer Ring Road, Marathahalli, Bangalore-560037, India}
\email{}
%%%\collaboration{all}{The Terra Mater collaboration}

\begin{abstract}
 We present a broadband spectral and timing study of the bright atoll neutron star low-mass X-ray binary GX~9+9, using simultaneous NICER, NuSTAR, and MAXI observations from 2019 and 2022. The broadband spectra are well described by a combination of multicolor disk emission, thermal Comptonization of blackbody seed photons from the neutron star surface or boundary layer, and associated disk reflection. We report a clear detection of reflection features in this source, including a moderately broad Fe K$\alpha$ complex and a reflection hump, confirming an earlier detection from a different observation. These features likely arise from the illumination of the moderately ionized disk by thermally Comptonized emission from the boundary layer or warm corona. While the source remained in the soft spectral state during both observations, the 2022 observation exhibits a relatively larger thermal contribution, together with a steeper Comptonized component and lower coronal optical depth. Both reflection and disk-continuum modeling favor a truncated disk, with the latter indicating a smaller inner disk radius in 2022. The inferred radial extent of the boundary layer of $\sim8~r_{\rm g}$ may play a role in the observed disk truncation. Assuming that the disk is truncated at the magnetospheric radius, we obtain an upper limit on the magnetic field strength of $\lesssim3\times10^{10}$ G. Timing analysis reveals a relatively stable disk component and a highly variable Comptonized component, consistent with the typical characteristics of the soft state.
\end{abstract}

%\keywords{\uat{Galaxies}{573} --- \uat{Cosmology}{343} --- \uat{High Energy astrophysics}{739} --- \uat{Interstellar medium}{847} --- \uat{Stellar astronomy}{1583} --- \uat{Solar physics}{1476}}


\section{Introduction}

Low-mass X-ray binaries (LMXBs) hosting weakly magnetized neutron stars (NSs) accrete matter through the Roche-lobe filling mechanism from their companion star. With luminosities ranging from $\sim 10^{35}$ to $\sim10^{38}~{\rm erg~s^{-1}}$, these systems are among the brightest X-ray sources in the sky and exhibit distinct spectral and timing behaviors that are widely interpreted in terms of variations in the mass accretion rate. Based on the tracks NS-LMXBs trace in the X-ray color–color diagram (CCD) and hardness–intensity diagram (HID), they are broadly classified into Z and atoll sources \citep{Hasinger1989A&A...225...79H, vanderKlis1989ARA&A..27..517V}. As the name suggests, Z sources follow a characteristic Z-shaped track in the CCD and typically accrete near the Eddington limit ($\gtrsim 0.5~L_{\rm Edd}$, $L_{\rm Edd}$ being the Eddington luminosity; \citealt{vanderKlis2005ASIB..210..283V}), making them persistently luminous systems.

In contrast, atoll sources are relatively less luminous and exhibit a C-shaped track comprising two distinct branches: the island and banana states. At luminosities of $\sim 0.001$–$0.1~L_{\rm Edd}$, they reside in the island state, characterized by relatively hard spectra and strong aperiodic variability. As the luminosity increases to $\sim 0.1$–$0.5~L_{\rm Edd}$, they evolve into the banana branch, where the spectra become softer, and the variability is significantly reduced \citep{Done2007A&ARv..15....1D}. The atoll sources possess weaker magnetic fields than the Z sources \citep{Done2003MNRAS.342.1041D}. Nevertheless, the distinction between Z and atoll sources is not always clear-cut; transitional behavior between these classes has been observed in sources such as XTE~J1701--462 \citep{Lin2009ApJ...696.1257L}, likely driven by changes in the mass accretion rate. Furthermore, \citet{Done2003MNRAS.342.1041D} showed that atoll sources can occasionally trace Z-like patterns, whereas the reverse is not observed.

The broadband X-ray spectra of NS-LMXBs, particularly in the soft state, are complex and typically require multiple emission components. These include a multicolor disk blackbody (MCD) emission (with characteristic temperatures $kT \sim 1$ keV) arising from an optically thick, geometrically thin accretion disk, and a high-energy component attributed primarily to thermal Comptonization of soft seed photons in a hot electron population with temperatures of a few keV \citep{Barret2001AdSpR..28..307B, Paizis2006A&A...459..187P, Farinelli2008ApJ...680..602F} or bulk-motion Comptonization in a converging inflow close to the NS surface \citep{Titarchuk1998ApJ...493..863T}. In addition, a boundary layer (BL) that forms at the interface between the inner disk and the NS surface due to the difference in their angular velocities \citep{Shakura1988AdSpR...8b.135S, Inogamov1999AstL...25..269I} can contribute significantly to both the thermal emission and Comptonization process. The presence of the aforementioned components implies that seed photons from both the disk and the NS/BL can contribute to the Comptonized emission, as done by \cite{Lin2007ApJ...667.1073L}. However, the prevailing interpretation is that the Comptonized emission arises from inverse Compton scattering of blackbody seed photons from the NS surface in the BL or warm corona, while the soft thermal component is attributed to the MCD \citep{Pringle1977MNRAS.178..195P, Shakura1988AdSpR...8b.135S, Popham2001ApJ...547..355P}.

Another crucial component of the emission spectrum is the reflection component, produced by hard X-ray irradiation of the accretion disk from the corona or BL. The illuminating continuum in NS-LMXBs may be in the form of either a thermally Comptonized component from the corona or a blackbody emission spectrum from the NS surface or BL. The reflection spectrum is characterized by prominent spectral features, including the fluorescent Fe $K_\alpha$ line at $\sim 6$–$7$ keV and a reflection hump peaking at $\sim 20$–$30$ keV due to Compton scattering of the hard X-rays. If the reflecting material lies close to the NS, relativistic effects such as Doppler broadening and gravitational redshift can distort the iron line profile \citep{Fabian1989MNRAS.238..729F}. Modeling this feature provides constraints on the inner disk radius and, consequently, an upper limit on the NS radius, as the disk is expected to truncate before reaching the stellar surface \citep{Cackett2008ApJ...674..415C, Cackett2009ApJ...694L..21C, Chiang2016ApJ...831...45C}.

Variability studies provide key insights into the geometry and dynamics of the inner accretion flow. The most common approaches include the analysis of power density spectra (PDSs), fractional rms variability as a function of photon energy, and time lags as a function of frequency and energy. In the island (hard) state, PDSs are typically characterized by strong band-limited noise (BLN), whereas in the banana (soft) state, the BLN component weakens or disappears, and the variability is dominated by very-low-frequency noise (VLFN) with a power-law–like shape \citep{vanderKlis2006csxs.book...39V, Wijnands2017MNRAS.472..559W}. The rms–energy spectrum provides a means to probe the energy dependence of variability and to distinguish variable and stable spectral components. 

The bright atoll NS-LMXB GX~9+9 was discovered near the Galactic bulge during a sounding rocket experiment in 1967 \citep{Bradt1968ApJ...152.1005B}. The distance to the source remains uncertain, with estimates in the range of $\sim 5$–$10$ kpc \citep{Christian1997ApJS..109..177C, Savolainen2009MNRAS.393..569S}. Optical and X-ray observations have revealed a $\sim 4.2$ hr orbital modulation in the light curves \citep{Hertz1988ApJ...331..764H, Schaefer1990ApJ...354..720S, Harris2009ApJ...696.1987H}. The source remained persistently in the soft spectral state, with an average X-ray luminosity of $\approx 230$ mCrab in the $2–20$ keV band \citep{Iaria2020A&A...635A.209I}. Previous studies using RXTE and INTEGRAL have suggested that the broadband spectra can be modeled both with and without reflection components \citep{Kong2006MNRAS.368..781K, Savolainen2009MNRAS.393..569S, Gogus2007ApJ...659..580G}.

Based on XMM-Newton and BeppoSAX observations, \citet{Iaria2020A&A...635A.209I} showed that reflection models are required to adequately describe the broadband spectrum of GX~9+9, even without the prominent presence of reflection features. The authors incorporated the convolution model \texttt{RFXCONV} \citep{Done2006MNRAS.367..659D, Kolehmainen2011MNRAS.416..311K} to account for reflection from an ionized disk, and the thermal Comptonized model \texttt{nthcomp} \citep{Zdziarski1996MNRAS.283..193Z} for the continuum, where the seed photons can be either from the disk or NS/BL. Their analysis yielded an inclination angle of $\sim 39\degr$–$60\degr$, consistent with the upper limit of $\sim 70\degr$ reported for this source \citep{Schaefer1990ApJ...354..720S, Savolainen2009MNRAS.393..569S}. Subsequently, \citet{Chatterjee2023MNRAS.521L..74C} showed that the $0.7$–$25$ keV spectrum from AstroSat observations can be described by Comptonized emission from an optically thick BL, together with a thermal component modeled either as an MCD or a single-temperature blackbody. The first clear detection of a broad Fe $K_\alpha$ line in GX~9+9 was reported by \citet{Ursini2023A&A...676A..20U} using simultaneous NuSTAR and IXPE observations obtained in 2022. The authors modeled the continuum assuming bulk-motion Comptonization and described the reflection spectrum using the \texttt{relxillNs} \citep{Garcia2022ApJ...926...13G} model, specifically tailored for NS-LMXBs, considering a blackbody input spectrum. Notably, polarization measurements in both \citet{Chatterjee2023MNRAS.521L..74C} and \citet{Ursini2023A&A...676A..20U} are also consistent with a Comptonizing region associated with the BL.

 In this work, we present a broadband spectral and timing study of GX~9+9 in the soft state using two sets of simultaneous NICER \citep{Gendreau2016SPIE.9905E..1HG}, NuSTAR \citep{Harrison2013ApJ...770..103H}, and MAXI \citep{Matsuoka2009PASJ...61..999M} observations obtained in 2019 and 2022. We show that thermal Comptonization of blackbody seed photons from the NS surface by thermal electrons in the warm corona or BL can describe the high-energy emission, providing an alternative to the bulk Comptonization scenario proposed by \citet{Ursini2023A&A...676A..20U}. We report the second clear detection of a moderately broad Fe $K_\alpha$ line, accompanied by a reflection hump, in the 2019 observation, following the first such detection reported by \citet{Ursini2023A&A...676A..20U} in the 2022 observation. Using both phenomenological and physically motivated models, we investigate the disk-corona geometry and the properties of the reflection features and compare them between the two observations. We also examine the roles of the BL and NS magnetic field in determining the disk truncation in the soft spectral state. Finally, we investigate the variability properties in terms of PDS and rms–energy spectra. This paper is organized as follows: Section~\ref{sec:obs} describes the observations and data reduction; Section~\ref{sec:analysis} presents the spectral and timing results; Section~\ref{sec:discussion} discusses the implications of our findings; and Section~\ref{sec:conclusion} summarizes our conclusions.

\section{Observations and data reduction}
\begin{figure}
    \centering
    \includegraphics[width=\columnwidth]{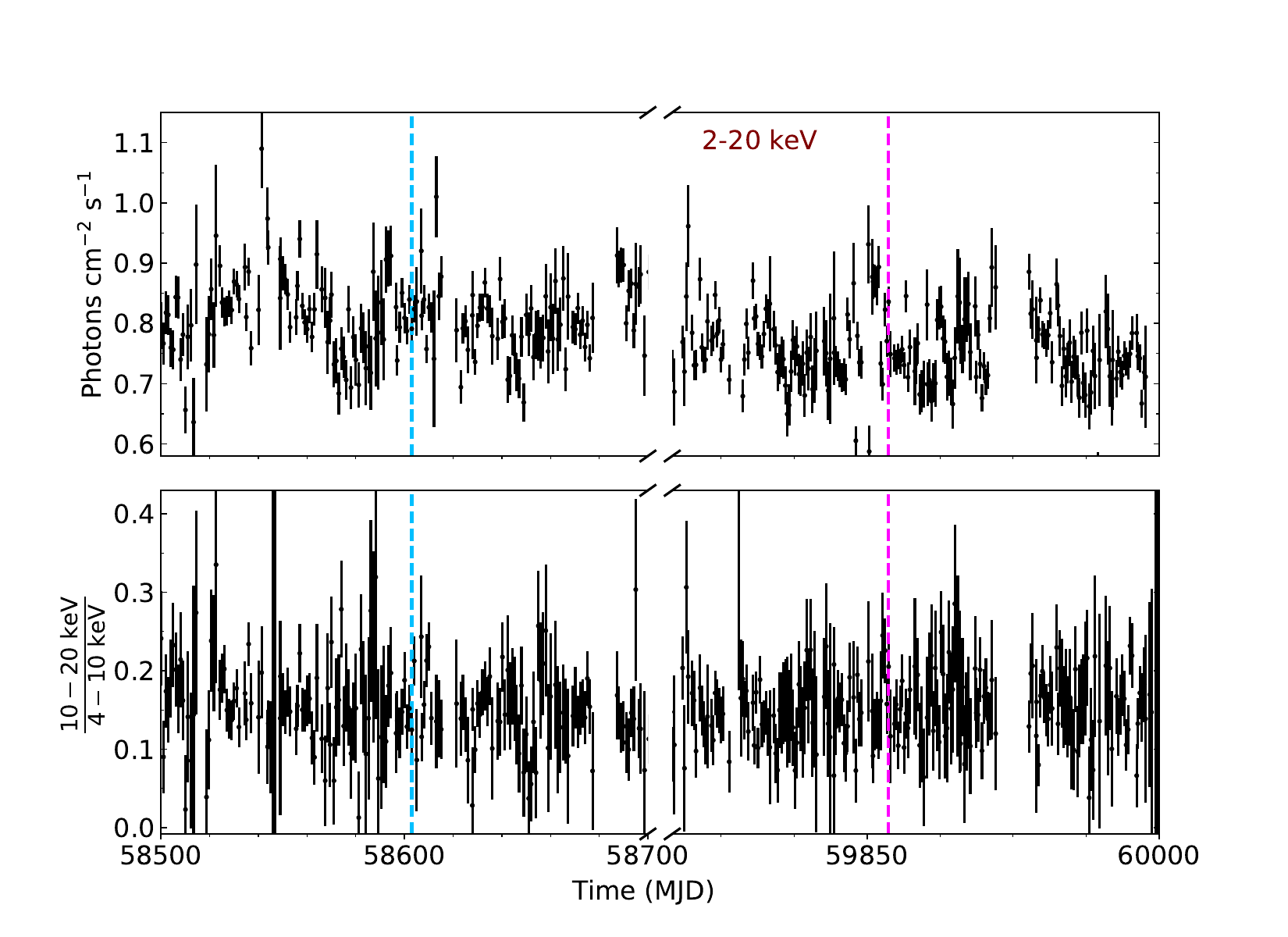}
    \caption{Top panel: MAXI long-term light curve in the 2--20 keV band. Bottom panel: HR between the 10--20 keV and 4--10 keV bands. Light curves were generated using the MAXI on-demand service for MJD 58500--60000.}
    \label{fig:maxi_ltcrv_hr}
\end{figure}

{\label{sec:obs}}
We analyze two sets of observations of GX~9+9 obtained in 2019 and 2022. The first dataset corresponds to a simultaneous NICER and NuSTAR observation carried out in April 2019. The second dataset consists of a NuSTAR observation performed in October 2022. In addition, we utilize MAXI on-demand data (see Section \ref{sec:maxi_data_reduction}) covering the same time interval as the 2022 NuSTAR observation. The details of these observations are summarized in Table~\ref{tab:obs_ids}. The positions of these observations on the long-term MAXI light curve, derived using the MAXI on-demand data in the interval of MJD 58500 and MJD 60000, are shown in Figure~\ref{fig:maxi_ltcrv_hr}. We also derived the hardness ratio (HR) by dividing the count rate in the $10–20$ keV band by that in the $4–10$ keV band. During both observations, the MAXI count rate remains similar, suggesting a persistent behavior of the source. However, in contrast to the $\sim$1--2 month HR variations reported by \citet{Ursini2023A&A...676A..20U} using MAXI auto-generated data (their Figure~1), we find no significant HR variations in the MAXI on-demand data. \citet{Corbet2024ApJ...976..137C} reported a similar modulation in the automatically generated MAXI light curve of the high-mass X-ray binary XTE~J1829--098, where this issue was attributed to the precession period of the International Space Station (ISS).
 %%% MAXI light curve and HR

\subsection{NICER}
We used the standard task \texttt{nicerl2} within HEASoft (v.6.36) to process the NICER observation along with the recent CALDB (v.xti\_20240206). The time-averaged source and background spectra were extracted using the standard script \texttt{nicerl3-spec}, where we used the `3C50' model for the background. We then used the \texttt{nicerl3-lc} script for generating the light curve.

\subsection{NuSTAR}
We processed the NuSTAR data using the standard task \texttt{nupipeline} of the \texttt{NuSTARDAS} package, along with the latest calibration files. Since the source count rate was much higher than 100 counts/s, we opted for the condition \texttt{statusexpr=``(STATUS==b0000xxx00xxxx000)\&\& \hspace{1cm}(SHIELD==0)''}, while using the \texttt{nupipeline} task. We considered a circular region of $100 \arcsec$ centered on the source position and an off-source region of the same radius to extract the source and background spectra, respectively, using the standard task \texttt{nuproducts}. This task also generated the response matrix files (RMFs) and ancillary response files (ARFs).

\subsection{MAXI}\label{sec:maxi_data_reduction}
 We used the MAXI on-demand service{\footnote{A web interface (\url{https://maxi.riken.jp/mxondem}) that allows users to derive calibrated source and background light curves, spectra, and response files for the MAXI instruments.}, which follows the procedure mentioned in \cite{Nakahira2012PASJ...64...13N}, to extract time-averaged source and background spectra from the onboard Gas Slit Camera (GSC), covering the same time interval as the 2022 NuSTAR observation. The radii of the source and background regions were kept at their default values of $1\degr.6$ and $3\degr$, respectively. The MAXI/GSC data provide complementary coverage down to $\sim 2$ keV, extending the broadband spectral constraints below the NuSTAR bandpass.

%%% Observation details
\begin{table*}
	\caption{List of Observations of GX~9+9}
 %  { log of observations}
   \centering
   \begin{tabular}{lcccccc}
      \hline
      \hline
    % & \multicolumn{2}{c}{$0.7-3$ keV} & \multicolumn{2}{c}{$3-10$ keV} 
    % \\\hline
    % & Epoch - 1 & Epoch - 2 & Epoch - 1 & Epoch - 2 \\
    Obs No & Instrument & Obs ID & Start Time & Stop Time & Eff. Exp.(s) \\
    &  & & (yyyy-mm-dd) & (yyyy-mm-dd) & \\
     \hline        
         Obs1  & NICER & $2050490101$ & 2019-04-30 09:24:13 & 2019-04-30 23:33:43 & 16910 \\
           & NuSTAR/FPMA & $30401015002$ & 2019-04-30 09:26:09 & 2019-05-01 09:16:09 & 30230 \\
           &    FPMB     & ...           & 2019-04-30 09:26:09 & 2019-05-01 09:16:09 & 30670 \\ \\

        Obs2    & MAXI &    ... &  2022-10-09 10:21:09  &   2022-10-10 14:51:09 & 2638  \\
                & NuSTAR/FPMA & 30801021002  &   2022-10-09 10:21:09 & 2022-10-10 14:51:09 & 39140 \\
                &   FPMB      & ...          &  2022-10-09 10:21:09  &  2022-10-10 14:51:09 & 39680 \\   
     \hline
   \end{tabular}
   \label{tab:obs_ids}
\end{table*}     

\section{Analysis and Results}{\label{sec:analysis}}
We constructed CCD by deriving light curves in the 3--6 keV, 6--10 keV, and 10--30 keV bands from NuSTAR observations. The soft and hard colors were defined as the count-rate ratios between the 6--10 keV and 3--6 keV bands, and the 10--30 keV and 6--10 keV bands, respectively. The resulting CCD is shown in Figure~\ref{fig:CCD_HID_nustar}, where both observations trace a similar track. This behavior indicates that the source remained in the soft state or the banana branch during the two observations analyzed here.
\begin{figure}
    \centering
    \includegraphics[width=\columnwidth]{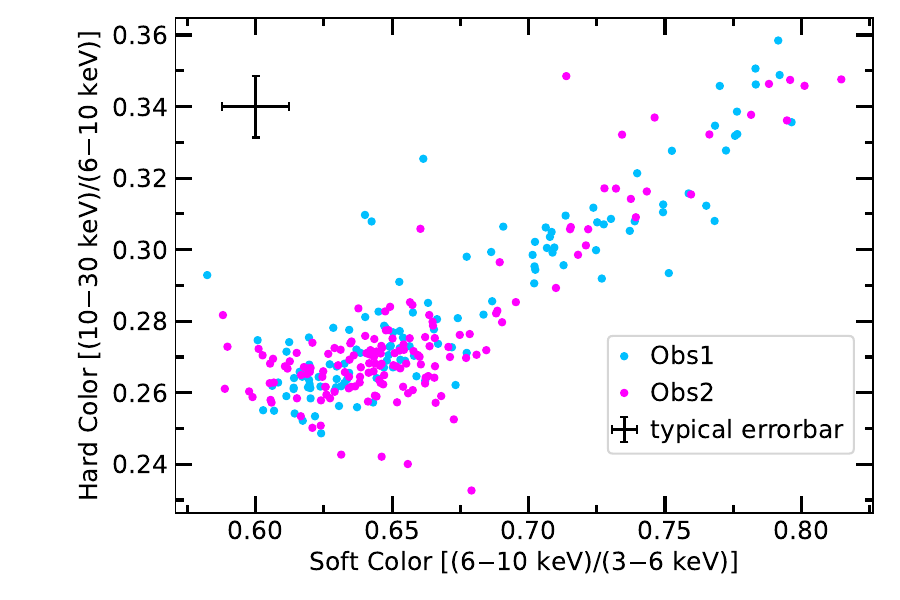}
    \caption{CCD derived from NuSTAR data for both observations using 400 s time bins.}
    \label{fig:CCD_HID_nustar}
\end{figure}

\subsection{Broadband Spectral Analysis}
We performed broadband spectral analysis using the X-ray spectral fitting package XSPEC \citep[v.12.15.1;][]{Arnaud1996ASPC..101...17A}, and quoted all the uncertainties in the spectral parameters at the $90\%$ confidence ($\Delta \chi^2 \approx2.71$) level. We applied the optimal binning algorithm \citep{Kaastra2016A&A...587A.151K}, with the additional requirement of 25 counts per bin, to all spectral data to use the $\chi^2$ statistic. To account for any calibration uncertainties and differences in relative normalizations between different instruments, we used the multiplicative component \texttt{plabs} in all our models. We fixed the parameters $\Delta \Gamma$ and $K$ at 0 and 1, respectively, for NuSTAR/FPMA data, and kept them free to vary for all other instruments. However, we did not observe any significant differences in spectral hardness between MAXI and NuSTAR observations, so we fixed $\Delta \Gamma$ at 0 for the MAXI observation.
%%% NuSTAR_resilduals
\begin{figure}
    \centering
    \includegraphics[width=\columnwidth]{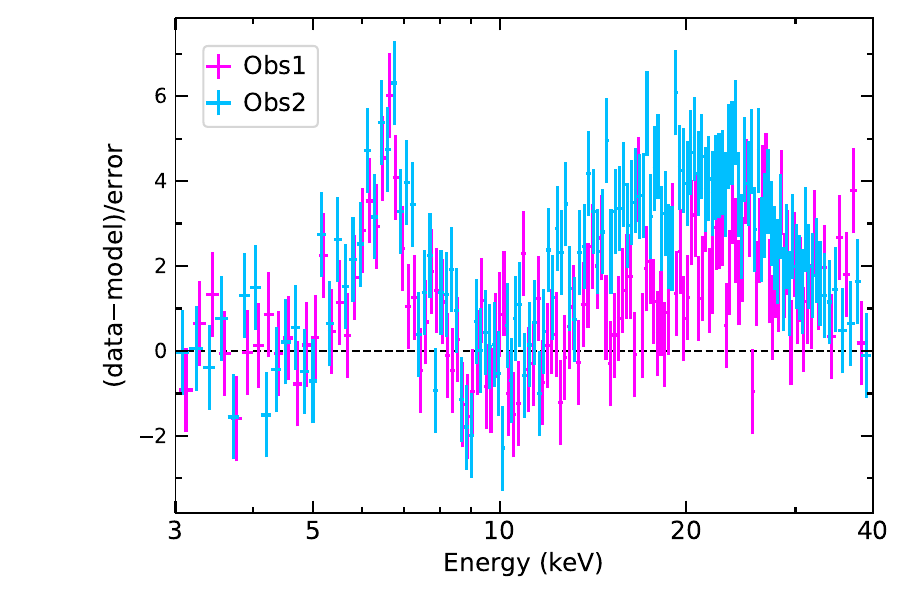}
    \caption{NuSTAR/FPMA data residuals with respect to the fitted model \texttt{tbabs(thcomp*bbody+diskbb)}. Spectral fitting is performed in the 3--5 keV and 9--12 keV bands.}
    \label{fig:fpma_res}
\end{figure}

We started by jointly fitting the NICER and NuSTAR/FPMA and FPMB spectral data in the 0.7--40 keV band for Obs1, and the MAXI/GSC and NuSTAR data in the 2--40 keV band for Obs2. NuSTAR data above 40 keV were excluded, as the background becomes comparable to the source, and NICER data below 0.7 keV were omitted to avoid calibration uncertainties. As an initial model, we employed an MCD component, \texttt{diskbb} \citep{Mitsuda1984PASJ...36..741M}, to describe the thermal emission from the accretion disk, a single-temperature \texttt{bbody} to account for the thermal emission from the NS surface, and \texttt{thcomp} for the thermal Comptonization. Here, we convolved \texttt{thcomp} with \texttt{bbody}, assuming the seed photons for Comptonization are from the NS surface. Since \texttt{thcomp} is a convolution model, the energy grid was extended from 0.01 to 1000 keV using the \texttt{energies} command in XSPEC to ensure accurate treatment of the Comptonized spectrum. Galactic absorption was modeled using \texttt{tbabs}, with abundances and cross sections from \citet{Wilms2000ApJ...542..914W} and \citet{Verner1996ApJ...465..487V}, respectively. This model, \texttt{plabs*tbabs(thcomp*bbody+diskbb)}, could not describe the spectral data, and returned a poor spectral fit, with $\chi^2/{\rm dof} = 447.7/416$ and $333.3/262$ for Obs1 and Obs2, respectively, leaving prominent residuals at the Fe K emission band ($\sim 6-7$ keV) and reflection hump region ($\sim15-30$ keV). Figure~\ref{fig:fpma_res} shows the NuSTAR/FPMA data residuals to the model \texttt{tbabs(thcomp*bbody+diskbb)}, where the presence of a fluorescent Fe $K_\alpha$ line and reflection hump is clearly observed.

The addition of a \texttt{gaussian} component significantly improves the spectral fit, with $\Delta\chi^2 \approx -54$ and $-70$ for Obs1 and Obs2, respectively, for three additional degrees of freedom. 
%Joint spectral fitting using NICER/MAXI, and NuSTAR spectral data using models M1 and M2
\begin{figure*}
    \centering
    \includegraphics[width=0.48\textwidth]{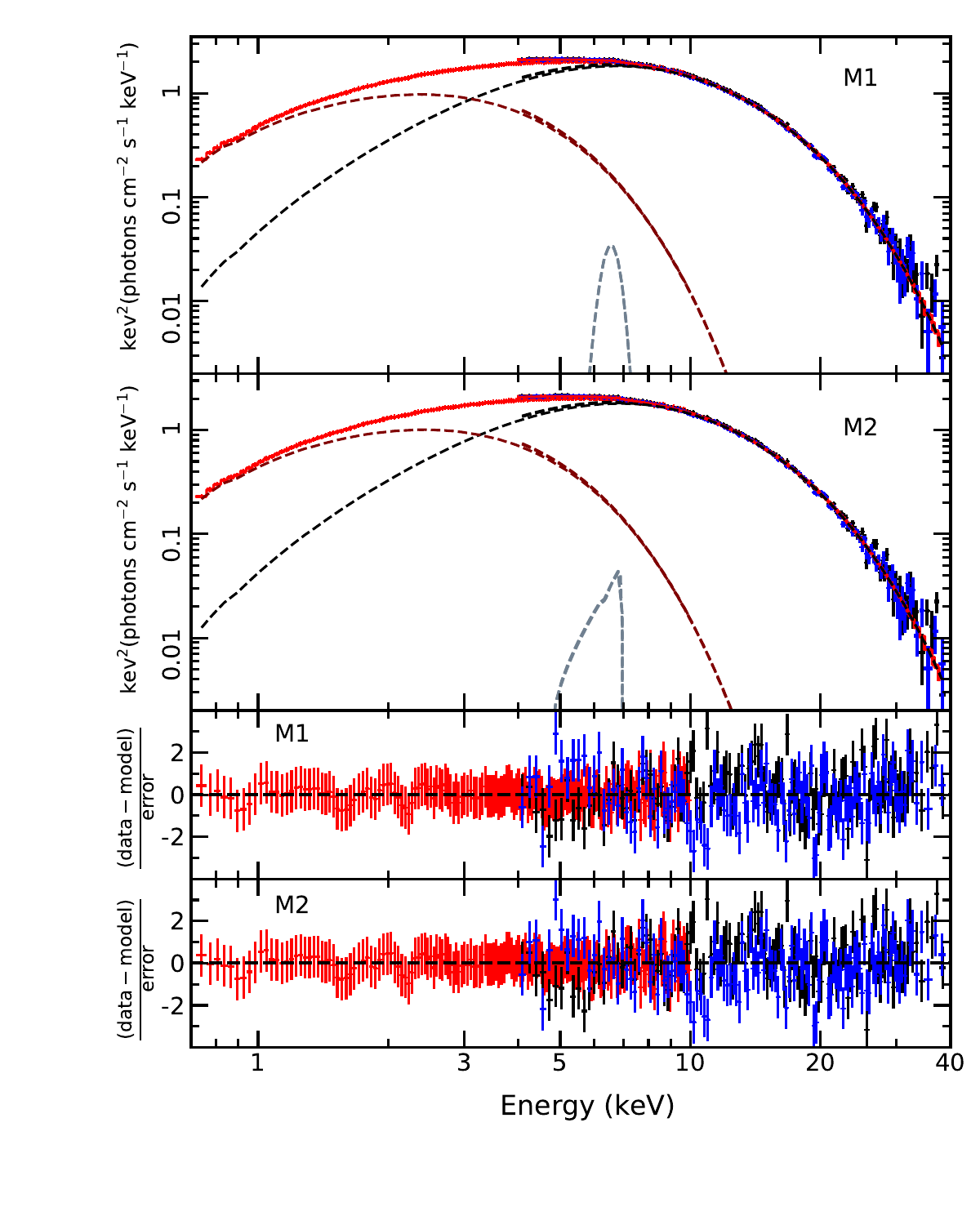}
    \includegraphics[width=0.48\textwidth]{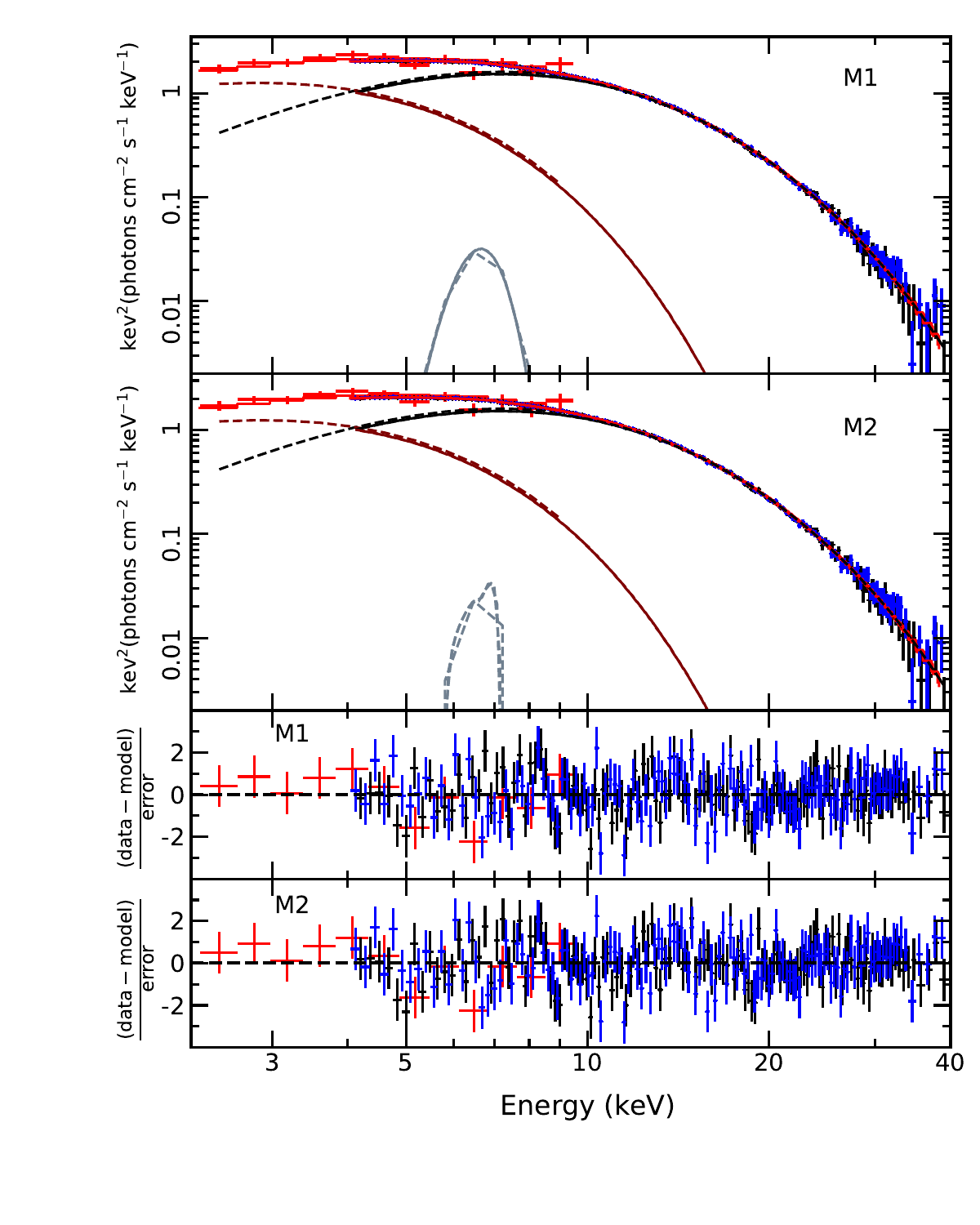}
    \caption{Left panel: joint NICER (red), NuSTAR/FPMA (black), and FPMB (blue) spectral data for Obs1. Right panel: joint MAXI (red), NuSTAR/FPMA (black), and FPMB (blue) spectra for Obs2. The spectra are fitted with model M1, \texttt{plabs*tbabs(diskbb+thcomp*bbody+gaussian)}, and model M2, \texttt{plabs*tbabs(diskbb+thcomp*bbody+diskline)}. The individual model components are shown as dotted lines: disk emission (maroon), thermal Comptonization (black), and line emission (gray).}
    \label{fig:m1_m2}
\end{figure*}
 The covering fraction parameter ($C_{\rm f}$) within \texttt{thcomp} was initially allowed to vary freely but was found to peg at the upper limit, yielding $C_{\rm f} > 0.9$ for Obs1. We therefore fixed it at 1 for both observations. The resulting model, \texttt{plabs*tbabs(diskbb+thcomp*bbody+gaussian)} (hereafter M1), provides an adequate fit to the spectra, with $\chi^2/{\rm dof} = 393.8/413$ and $263.6/260$ for Obs1 and Obs2, respectively. The best-fit spectral parameters are listed in Table~\ref{tab:M1_M2}, and the best-fit model with spectral data is shown in Figure~\ref{fig:m1_m2}.

%%% Best-fit Spectral Parameters with phenomenological models
\begin{table*}[!t]
\caption{Best-fit Broadband X-Ray Spectral Parameters Derived Using Models M1 and M2}
\centering
\small
\setlength{\tabcolsep}{5pt}
\renewcommand{\arraystretch}{0.92}

\begin{tabular}{lccccccc}
\hline
\hline

& & \multicolumn{2}{c}{Obs1} && \multicolumn{2}{c}{Obs2} \\
\cline{3-4} \cline{6-7}

Component & Parameter & M1 & M2 && M1 & M2 \\
\hline

\texttt{tbabs} & $N_{\rm H}~[10^{22}~{\rm cm^{-2}}]$ & $0.22\pm0.01$ & $0.22\pm0.01$ & & $0.2^f$ & $0.2^f$ \\

\texttt{diskbb} & $kT_{\rm in}$ (keV) & $0.91\pm0.03$ & $0.93\pm0.03$ & & $1.16^{+0.11}_{-0.12}$
& $1.18^{+0.12}_{-0.08}$ \\
& $N_{\rm disk}$ & $257.8^{+31.7}_{-30.0}$ & $242.9^{+27.2}_{-21.7}$ & & $99.2^{+38.6}_{-25.0}$
& $92.6^{+54.8}_{-24.5}$ \\

\texttt{thcomp} & $\Gamma$ & $2.62^{+0.14}_{-0.13}$ & $2.72^{+0.15}_{-0.13}$ & & $3.28^{+0.62}_{-0.43}$ & $3.22^{+0.67}_{-0.42}$ \\
& $kT_{\rm e}$ (keV) & $3.0\pm0.1$ & $3.1\pm0.1$ & & $3.43^{+0.56}_{-0.29}$ & $3.38^{+0.30}_{-0.16}$ \\
& $C_{\rm f}$ & $1^f$ & $1^f$ & & $1^f$ & $1^f$ \\

\texttt{bbody} & $kT_{\rm bb}$ (keV) & $1.33\pm0.06$ & $1.37^{+0.06}_{-0.08}$ & & $1.59^{+0.14}_{-0.12}$
& $1.58^{+0.16}_{-0.19}$ \\
& $N_{\rm bb}$ ($10^{-2}$) & $4.50^{+0.04}_{-0.05}$ & $4.47\pm0.04$ & & $3.76^{+0.34}_{-0.26}$
& $3.83^{+0.33}_{-0.15}$ \\

\texttt{gaussian} & $E_{\rm line}$ (keV) & $6.5\pm0.1$ & ... & & $6.56^{+0.11}_{-0.23}$ & ... \\
& $\sigma$ (keV) & $0.29^{+0.14}_{-0.11}$ & ... & & $0.54^{+0.46}_{-0.16}$ & ... \\
& norm ($10^{-4}$) & $6.1^{+2.7}_{-1.9}$ & ... & & $9.9^{+16.0}_{-3.4}$ & ... \\

\texttt{diskline} & $E_{\rm line}$ (keV) & ... & $6.37^{+0.21}_{-0.08}$ & & ... & $6.64^{+0.10}_{-0.11}$ \\
& $R_{\rm in}$ ($r_{\rm g}$) & ... & $<89$ & & ... & $40.7^{+29.9}_{-17.7}$ \\
& norm ($10^{-4}$) & ... & $10.2^{+3.4}_{-5.1}$ & & ... & $6.19^{+1.8}_{-1.4}$ \\

\hline

Cross-calibration & $\Delta \Gamma_{\rm NICER}$ & $-0.09\pm0.01$ & $-0.097\pm0.011$ & & $0^f$ & $0^f$ \\
& $K_{\rm NICER/MAXI}$ & $0.81\pm0.02$ & $0.80\pm0.02$ & & $1.06\pm0.05$ & $1.06\pm0.05$ \\
& $K_{\rm FPMB}$ & $0.982\pm0.001$ & $0.982\pm0.001$ & & $1.005^{+0.001}_{-0.005}$ & $1.005\pm0.001$ \\

\hline

$\chi^2/{\rm dof}$ & & $393.8/413$ & $389.6/413$ & & $263.6/260$ & $270.6/260$ \\
AIC & & 419.8 & 415.6 & & 285.6 & 292.5 \\
BIC & & 472.5 & 468.3 & & 325 & 332 \\
\hline
\multicolumn{7}{c}{Derived Values} \\
\hline

${\rm Flux^{unabs}_{total}}$ & & $8.54\pm0.05$ & & & $7.60\pm0.03$ & \\
$F_{\rm disk}~ (\%)$ & & 44 & & & 52 & \\
$F_{\rm bbody}~ (\%)$ & & 44 & & & 41 & \\
$F_{\rm cor}~ (\%)$ & & 12 & & & 7 & \\
$F_{\rm gaussian}~ (\%)$ & & 0.08 & & & 0.13 & \\
& $\tau$ & $6.9\pm0.6$ & & & $4.8\pm1.3$ & \\

\hline
\end{tabular}
\label{tab:M1_M2}
\tablecomments{
$^f$ indicates fixed parameters. Flux is derived over the 0.001--500 keV band and is given in units of $10^{-9}$ erg cm$^{-2}$ s$^{-1}$.}
\end{table*}

The best-fit hydrogen column density, $N_{\rm H}$, is found to be $\approx 0.2 \times 10^{22}~{\rm cm^{-2}}$, consistent with previously reported measurements \citep{Iaria2020A&A...635A.209I, Chatterjee2023MNRAS.521L..74C} as well as the Galactic value along the line of sight. The best-fit photon index, $\Gamma$, and electron temperature, $kT_e$, indicate that the source was in the soft spectral state. The photon index increases from 
$2.62^{+0.14}_{-0.13}$ in Obs1 to $3.28^{+0.62}_{-0.43}$ in Obs2, indicating a steeper spectrum in the latter observation. This is accompanied by higher inner disk and seed photon temperatures (Table~\ref{tab:M1_M2}), suggesting a stronger contribution from thermal emission in Obs2. To understand the nature of the corona, we estimated the optical depth, $\tau$, of the Comptonizing region from the best-fit values of $\Gamma$ and $kT_{\rm e}$ using the relation
\begin{equation}
\tau = \sqrt{\frac{9}{4} + \frac{3m_e c^2}{kT_e \left[(\Gamma + \frac{1}{2})^2 - \frac{9}{4}\right]}} - \frac{3}{2},
\end{equation}
as given by \citet{Sunyaev1980A&A....86..121S}. Here, $m_e$ and $c$ denote the electron mass and the speed of light, respectively. The inferred values (see Table~\ref{tab:M1_M2}) indicate that the coronal flow is optically thick in both observations, with $\tau$ being higher for Obs1 than for Obs2.

 We derived the color-corrected inner disk radius from the \texttt{diskbb} normalization using the formula 
 $R_{\rm in} = f^2_{\rm col} (N_{\rm disk}/\cos i)^{1/2} D_{10}$, where $f_{\rm col}$ is the color-correction factor, $i$ is the inclination angle, and $D_{10}$ is the distance to the source in units of 10 kpc. We assumed $f_{\rm col}$, $i$ and distance to be 1.7 \citep{Shimura1995ApJ...445..780S}, $40\degr$, and 7.5 kpc respectively. It is worth noting that \citet{Kubota1998PASJ...50..667K} introduced an additional correction factor, $\xi=0.41$, to account for the inner-boundary condition, which is not incorporated into the \texttt{diskbb} model. Their derivation is based on a Newtonian disk with a zero-torque inner-boundary condition. However, if the disk is truncated well outside the ISCO due to evaporation (see below), the torque at the truncation radius may differ from that assumed in the standard zero-torque prescription. In this case, adopting $\xi=1$ may provide an alternative approximation for estimating the inner disk radius. The inferred disk radii of $19.2\pm1.2~r_{\rm g}$ and $11.9^{+2.1}_{-1.6}~r_{\rm g}$ for Obs1 and Obs2, respectively, suggest that the accretion disk remains truncated. The centroid energy of the Fe K$\alpha$ line is consistent with emission from a weakly or moderately ionized disk, consistent with a truncated accretion disk. We find the unabsorbed bolometric flux in the 0.001--500 keV band is moderately higher in Obs1 than in Obs2 (see Table~\ref{tab:M1_M2}). We further estimated the power supplied by the warm Comptonizing plasma as $F_{\rm cor}$ = $F_{\rm thcomp} - F_{\rm bbody}$, where $F_{\rm bbody}$ is the intrinsic flux of the blackbody component, and $F_{\rm thcomp}$ is the total flux of the \texttt{thcomp*bbody} component. The latter includes fluxes from both Comptonization and blackbody emission, and their difference represents the additional power supplied by the corona to express the equipartition of the source power into the intrinsic disk and corona components.
 
 We then replaced the \texttt{gaussian} component with a more physical and relativistic disk emission line model \texttt{diskline} \citep{Fabian1989MNRAS.238..729F} for a non-rotating compact object. This model provides an independent estimate of the inner disk radius by modeling the shape of the Fe $K_\alpha$ line profile. We used a fixed emissivity index of 3 and an inclination angle of $40\degr$ \citep{Iaria2020A&A...635A.209I}. This model, \texttt{plabs*tbabs(diskbb+thcomp*bbody+diskline)}, hereafter referred to as M2, provides almost similar fit statistics (see Table{}\ref{tab:M1_M2}). The best-fit model is shown in Figure~\ref{fig:m1_m2} and the best-fit parameters are given in Table~\ref{tab:M1_M2}. The inner disk radii of $\gtrsim 23~r_{\rm g}$ inferred from reflection modeling for both observations are in close agreement with those obtained from the disk continuum modeling. 

We also performed a statistical comparison between the models by deriving the Akaike information criterion, AIC= $-2 \ln \mathcal{L} + 2k$ \citep{Akaike1974ITAC...19..716A}, and the Bayesian information criterion, 
BIC=$ -2 \ln \mathcal{L} + k \ln(N)$ \citep{Schwarz1978AnSta...6..461S}. Here $k$ is the number of free parameters, $N$ is the total number of data points, and $\mathcal{L}$ is the likelihood of the best-fitting model, which can be calculated as $-2 \ln \mathcal{L}=\chi^2_{\rm min}$. The model with the lowest AIC and BIC values is statistically preferred. The values listed in Table~\ref{tab:M1_M2} indicate a slight preference for M2 over M1 in Obs1, while M1 is marginally favored in Obs2.

%Joint spectral fitting using NICER/MAXI, and NuSTAR spectral data using model M3
\begin{figure*}
    \centering
    \includegraphics[width=0.48\textwidth]{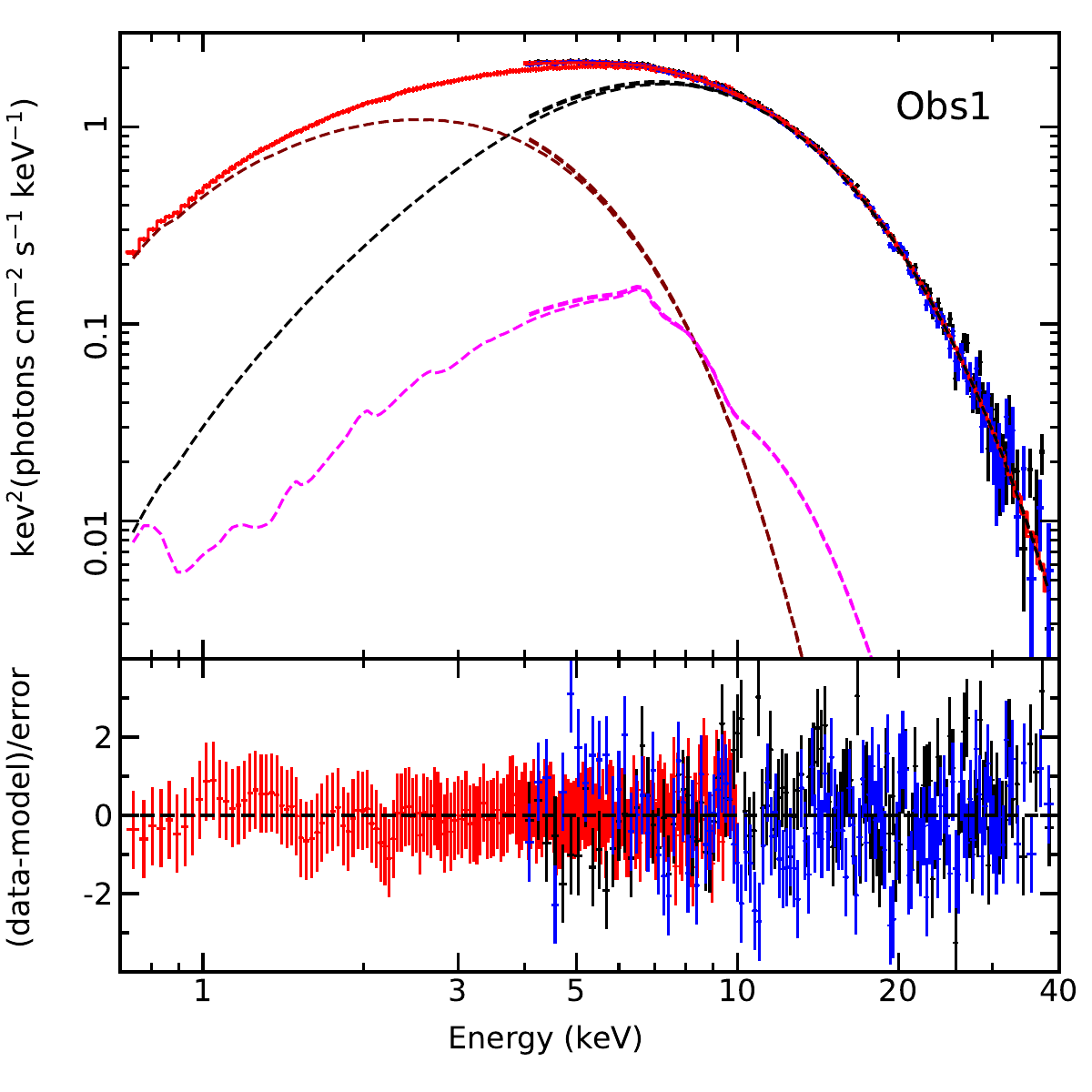}
    \includegraphics[width=0.48\textwidth]{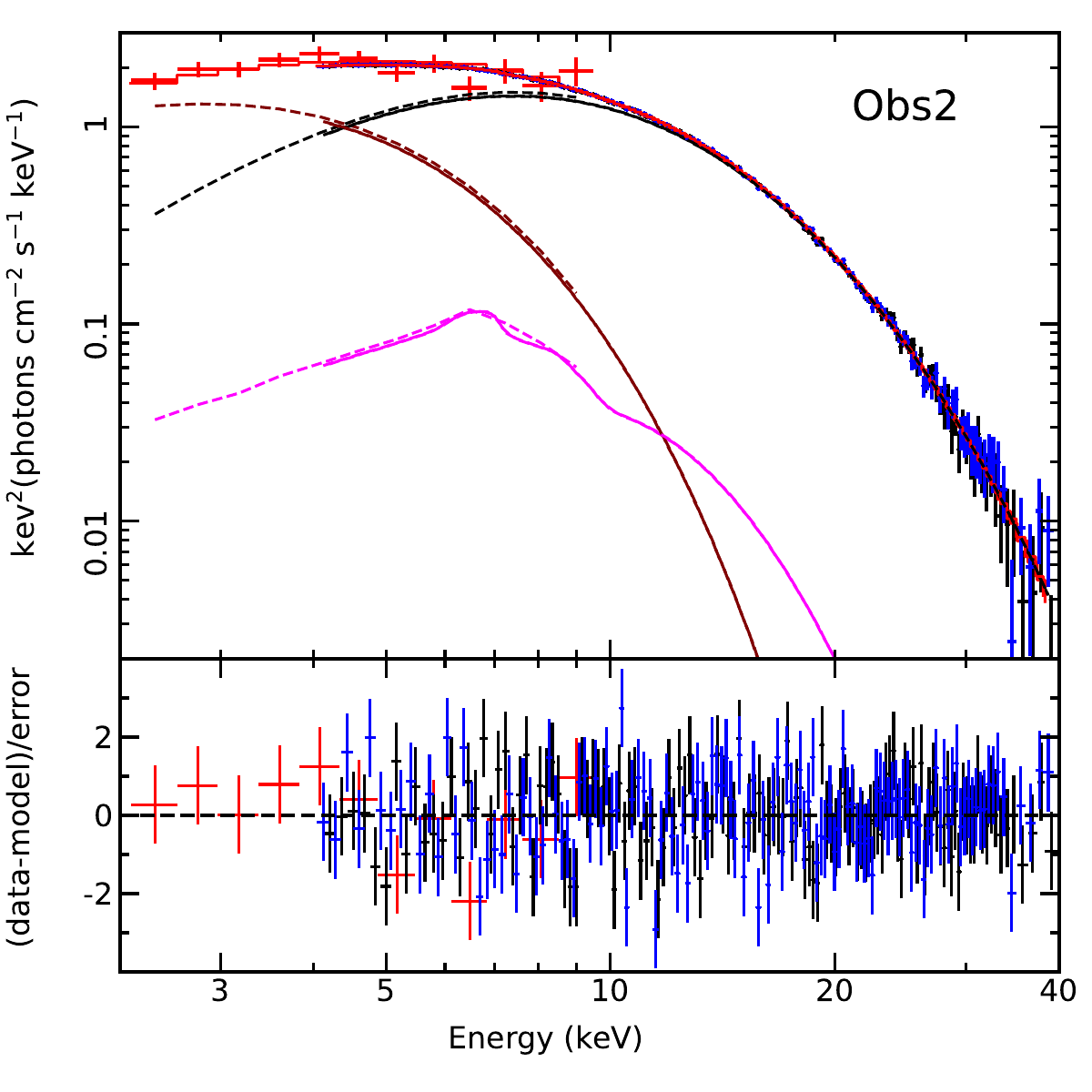}
    \caption{Left panel: joint NICER (red), NuSTAR/FPMA (black), and FPMB (blue) spectral data for Obs1. Right panel: joint MAXI (red), NuSTAR/FPMA (black), and FPMB (blue) spectral data for Obs2. The spectra are fitted with model M3, \texttt{plabs*tbabs(diskbb+thcomp*bbody+relxillNs)}. The individual model components are shown as dotted lines: disk emission (maroon), thermal Comptonization (black), and reflection component (magenta).}
    \label{fig:m3}
\end{figure*}

 Finally, we employed the relativistic reflection model \texttt{relxillNs}, which comes from the family of \texttt{relxill}. This model \texttt{relxillNs} has been specifically designed for NS-LMXBs, where the illuminating continuum is assumed to be a blackbody spectrum. This significantly differs from the Comptonized continuum (identical to \texttt{nthcomp}) in \texttt{relxillCp} or a cut-off powerlaw in \texttt{relxill}. We used \texttt{relxillNs} solely to describe the reflection component, fixing the reflection fraction, $R_{\rm f}$, to -1. The spin of the NS is fixed at 0.2 \citep{Ursini2023A&A...676A..20U}. The outer disk radius was fixed at 1000 $r_{\rm g}$, and a single emissivity profile ($\propto r^{-3}$) was used throughout the accretion disk. We tied the blackbody temperature of the illuminating continuum to the surface temperature of the NS. The density of the accretion disk, $N$, surrounding an NS is expected to be $\geq10^{20}~{\rm cm^{-3}}$ \citep{Shakura1973A&A....24..337S, Frank2002apa..book.....F, Ludlam2024Ap&SS.369...16L}. Thus we fixed $N$ at the highest model-defined value of $10^{19}~{\rm cm^{-3}}$. The remaining free parameters of \texttt{relxillNs} include the inner disk radius ($R_{\rm in}$), inclination angle ($i$), disk ionization ($\log \xi$), and normalization. This model, \texttt{plabs*tbabs(diskbb+thcomp*bbody+relxillNs)}, hereafter referred to as M3, provides a very good fit, with $\chi^2/{\rm dof} = 387.8/412$ and $252/259$ for Obs1 and Obs2, respectively. The best-fit spectral parameters along with the AIC and BIC values are listed in Table~\ref{tab:relxillns}, and the corresponding best-fit model and data are shown in Figure~\ref{fig:m3}. For Obs2, both the AIC and BIC favor model M3 over M1 and M2, where the reflection features are more prominent. For Obs1, M3 yields AIC and BIC values comparable to those of M1 and M2, which may be related to the relatively weaker reflection features in this observation. Notably, we also explored spectral fitting by allowing \texttt{relxillNs} to self-consistently calculate the illuminating continuum as the blackbody emission from the NS surface and the corresponding reflected component. However, this approach provided a poor spectral fit, suggesting that Comptonized emission from inverse Compton scattering of NS surface photons is more appropriate as the illuminating continuum.
  
 The best-fit parameters obtained with model M3 are consistent with those of the previous models within the uncertainties. The inclination angle remains poorly constrained; however, the upper and lower bounds are consistent with the values reported by \cite{Iaria2020A&A...635A.209I} and \cite{Ursini2023A&A...676A..20U}. The reflection modeling favors a moderately ionized and truncated accretion disk. We note that the relatively large uncertainties in $R_{\rm in}$ arise from the weak reflection features and limited data quality. Besides, the disk continuum modeling also favors a truncated disk having radii of $16.3^{+0.7}_{-0.8}~r_{\rm g}$ and $11.6^{+2.1}_{-1.5}~r_{\rm g}$ for Obs1 and Obs2, respectively.

%%% Reflection spectroscopy using relxillNs as the reflection component only.
\setlength{\tabcolsep}{1pt}
\renewcommand{\arraystretch}{1}
\begin{table}
\caption{Best-fit Broadband X-Ray Spectral Parameters Derived Using Model M3}
\centering
\begin{tabular}{lccc}
\hline
Component & Parameter & Obs1 & Obs2 \\
\hline

\texttt{tbabs} & $N_{\rm H}$ [$10^{22}~{\rm cm^{-2}}$] & $0.22\pm0.01$ & $0.2^{f}$ \\

\texttt{diskbb} & $kT_{\rm in}$ (keV) & $0.98^{+0.03}_{-0.04}$ & $1.2\pm0.1$ \\
               & $N_{\rm disk}$ & $208.7^{+18.2}_{-19.2}$ & $101.7^{+34.7}_{-26.5}$ \\

\texttt{thcomp} & $\Gamma$ & $3.03^{+0.33}_{-0.23}$ & $3.58^{+0.62}_{-0.44}$ \\
               & $kT_{\rm e}$ (keV) & $3.27^{+0.28}_{-0.22}$ & $3.69^{+0.66}_{-0.45}$ \\
               & $C_{\rm f}$ & $1^{f}$ & $1^{f}$ \\

\texttt{bbody} & $kT_{\rm bb}$ (keV) & $1.54^{+0.09}_{-0.13}$ & $1.67^{+0.12}_{-0.10}$ \\
              & $N_{\rm bb} (10^{-2})$ & $4.14^{+0.11}_{-0.27}$ & $3.56^{+0.18}_{-0.24}$ \\

\texttt{relxillNs} & $R_{\rm in}$ ($r_g$) & $56.8^{p}_{-47.5}$ & $43.4^{+79.7}_{-27.7}$ \\
                  & $i (\degr)$ & $20^{+14}_{p}$ & $37.6^{p}_{-14.5}$ \\
                  & $\log\xi$ & $3.28^{+0.50}_{-0.63}$ & $2.68^{+0.46}_{-1.13}$ \\
                  & $\log N$ & $19^{f}$ & $19^{f}$ \\
                  & $R_{\rm f}$ & $-1^{f}$ & $-1^{f}$ \\
                  & norm ($\times 10^{-4}$) & $3.2^{+2.5}_{-1.1}$ & $3.4^{+3.6}_{-1.0}$ \\
                  \hline
Cross-   &   $\Delta \Gamma _{\rm{NICER}}$   &   $-0.098\pm0.011$  &   $0^f$\\
calibration &   $K_{\rm{NICER}}$ &   $0.80\pm0.02$   &   \nodata \\
         &   $K_{\rm{MAXI}}$ &   \nodata   &   $1.05\pm0.05$\\
         &   $K_{\rm{FPMB}}$ &   $0.982\pm0.001$ &   $1.005\pm0.001$ \\
         \hline
$\chi^2$/dof & & $387.8/412$ & $252/259$ \\
AIC & & 415.8 & 276.2 \\
BIC & & 472 & 319.4 \\
\hline
                & & Derived values \\
                \hline
Flux$_{\rm tot}$ & & 8.67 & 7.87 \\

$F_{\rm disk}~ (\%)$ & & 48 & 52 \\
$F_{\rm bbody}~ (\%)$ & & 40 & 38.5 \\
$F_{\rm cor}~ (\%)$ & & $7.6$ & $5.6$  \\
$F_{\rm refl}~ (\%)$ & & 4.5   &   4.8    \\

                & $\tau$ & $5.4^{+0.8}_{-0.9}$ & $4.1\pm1.1$ \\
\hline

\end{tabular}

\tablecomments{
$^f$ indicates fixed parameters, $p-$ indicates that the parameter is pegged at the model-defined highest or lowest
bound during error estimation. Flux is derived over the 0.001--500 keV band and is given in units of $10^{-9}$ erg cm$^{-2}$ s$^{-1}$.
}
\label{tab:relxillns}
\end{table}

\subsection{Variability}
To investigate the variability properties of the source, we first extracted PDSs from the NICER and cross power density spectra \citep[CPDSs;][]{Bachetti2015ApJ...800..109B} from the NuSTAR data across different energy bands. The CPDS method derives the complex cross spectrum (cospectrum) from the FPMA and FPMB light curves via Fourier transforms. The real part of this cospectrum suppresses uncorrelated Poisson noise and mitigates the effects of data gaps, providing an alternative to the white-noise-subtracted PDS \citep[see][for more details]{Bachetti2015ApJ...800..109B}. We did not use the MAXI observation due to poor data quality. The \texttt{powspec} task within the \texttt{XRONOS} package was used to generate the Poisson-noise-subtracted and rms-normalized PDSs in the $0.5-3$ keV and $3-10$ keV bands, whereas the CPDSs were computed using the \texttt{HENDRICS} package \citep{Bachetti2015ApJ...800..109B} in the $3-10$ keV and $10-30$ keV bands. Figure~\ref{fig:pds} shows the resulting PDSs and CPDSs for both observations,
\begin{figure}
    \centering
    \includegraphics[width=0.9\columnwidth]{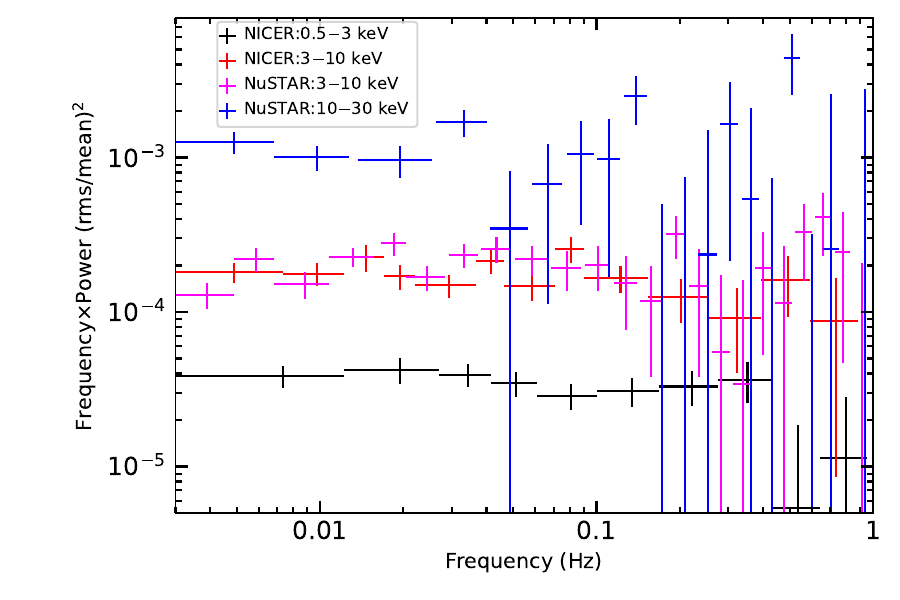}
    \includegraphics[width=0.9\columnwidth]{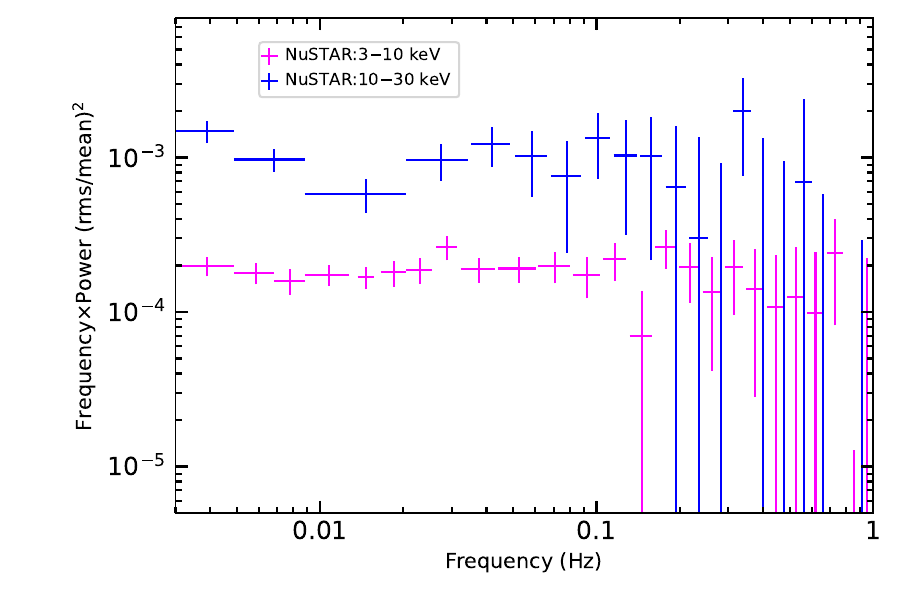}
    \caption{PDSs and CPDSs derived from the NICER and NuSTAR observations. The upper and lower panels correspond to Obs1 and Obs2, respectively. The plots are rebinned for clarity.}
    \label{fig:pds}
\end{figure}
and suggests that the variability power is lowest in the 0.5--3 keV band, dominated by disk emission, and increases with energy, reaching a maximum in the 10--30 keV band where thermal Comptonization dominates.

To investigate the origin of the energy-dependent variability, we derived the fractional rms amplitude as a function of photon energy for both observations. We employed the spectral-timing analysis package \texttt{Stingray} \citep{Huppenkothen2019JOSS....4.1393H, Huppenkothen2019ApJ...881...39H} to compute the fractional rms–energy spectra in the $0.5-10$ keV band over the $0.01-0.1$ Hz frequency range from the NICER data for Obs1. For the NuSTAR data, CPDSs were first extracted in different energy bands, and the fractional rms in the $0.01-1$ Hz range was then computed from each CPDS. The resulting fractional rms as a function of photon energy is shown in Figure~\ref{fig:rms_E}.
\begin{figure}
    \centering
    \includegraphics[width=0.48\textwidth]{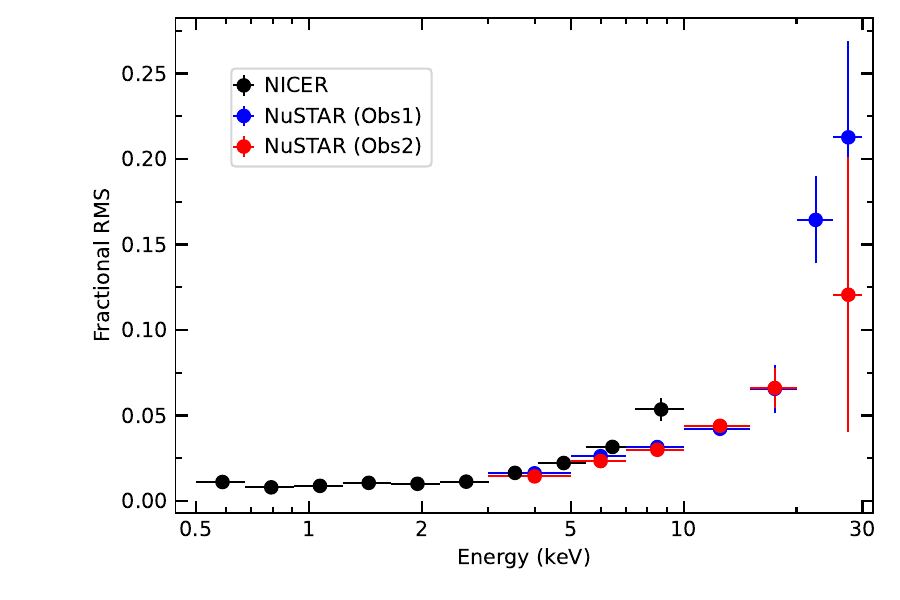}
    \caption{Fractional rms as a function of photon energy in the 0.01--0.1 Hz.}
    \label{fig:rms_E}
\end{figure}
 For Obs1, a little to moderate variability of $\lesssim 2\%$ is observed at soft energies, indicative of a stable or weakly variable accretion disk. The variability remains roughly constant up to $\sim 5$ keV and then gradually increases, reaching $\sim 22\%$ at higher energies, where the emission is typically dominated by thermal Comptonization. The variability pattern for Obs2 remains similar to that of Obs1 in the NuSTAR band.

\section{Discussion}{\label{sec:discussion}}
 We perform a broadband spectral and timing study of the bright atoll NS-LMXB GX~9+9 using NICER, NuSTAR, and MAXI observations obtained in 2019 and 2022. The long-term MAXI light curve and HR indicate that the source remained persistent during both observations. The variability in the HR reported by \cite{Ursini2023A&A...676A..20U} could be associated with the precession period of the ISS \citep{Corbet2024ApJ...976..137C}. The NuSTAR CCDs for both observations occupy a similar narrow banana-branch, with no evidence for transitions to the island state.

 The broadband spectra are well described by models consisting of MCD from the accretion disk, thermal Comptonization of seed photons from the NS surface in an optically thick warm flow or BL, and a reflection component produced by hard X-ray irradiation of the disk. Under this framework, all three spectral models provide comparable statistical fits to the broadband spectrum of Obs1, whereas model M3 is statistically preferred over M1 and M2 for Obs2, possibly due to the relatively strong reflection features. Modeling of the primary continuum with thermal Comptonization provides an alternative to the bulk-motion Comptonization scenario adopted by \citet{Ursini2023A&A...676A..20U}. Our choice of the illuminating continuum is also consistent with those of \citet{Iaria2020A&A...635A.209I} and \citet{Chatterjee2023MNRAS.521L..74C} for this source. Similar descriptions of the primary continuum have also been applied to several other NS-LMXBs, including Cyg~X-2 \citep{Ludlam2022ApJ...927..112L}, 4U~1702--429 \citep{Banerjee2024MNRAS.529.4311B}, and GX sources such as GX~3+1 \citep{Gnarini2024A&A...692A.123G}. Our spectral models provide an estimate of the additional power supplied by the corona through Comptonization to be in the range of $\sim6-12\%$ of the total flux. Such a small coronal contribution is consistent with the source being in the soft spectral state, in which the source spectrum is expected to be dominated by thermal emissions from the disk and NS surface. 
 
 The temperatures of the inner accretion disk, seed photons from the NS surface, and coronal temperature are in close agreement with those reported for the 2022 observation (our Obs2) by \citet{Ursini2023A&A...676A..20U} and for the XMM-Newton and BeppoSAX observations obtained in 2000--2001 by \citet{Iaria2020A&A...635A.209I}. The higher seed-photon temperature than that of the disk favors an origin of the seed photons for Comptonization in the NS surface or BL rather than the accretion disk. \citet{Chatterjee2023MNRAS.521L..74C} reported a relatively lower disk temperature and a higher coronal temperature based on AstroSat observations from 2020. They did not detect an iron emission line, and their study of flux-resolved spectra indicated spectral variability in the source. While the source remained persistently in the soft state, we observed a slight decrease in the source bolometric luminosity from $\sim 0.33~L_{\rm Edd}$ in Obs1 to $\sim 0.29~L_{\rm Edd}$. The spectral steepening, along with higher disk and seed-photon temperatures, suggests a relatively larger thermal contribution in Obs2. This is also in alignment with the lower coronal contribution and optical depth found in Obs2. The latter implies a lower scattering probability and, consequently, reduced Comptonization efficiency, consistent with our results. We note that Obs2 lacks low-energy coverage below 2 keV, which is important for constraining the disk temperature and normalization. On the other hand, for Obs1, the NICER exposure covers only the first $\sim17$ ks of the NuSTAR observation. Given the variable nature of GX~9+9, the non-simultaneity of the observations could introduce additional uncertainties in the inferred disk properties.

 To further investigate the origin of the seed photons, we estimated the apparent seed-photon emission radius, $R_{\rm s}$, from the XSPEC definition of the \texttt{bbody} normalization, defined as $=L/D^2_{10}$, where $L$ is the source luminosity in units of $10^{39}~\rm erg/s$ and $D_{10}$ is the source distance in units of 10 kpc. The luminosity is then related to $R_{\rm s}$ via $L = 4\pi R_{\rm s}^2 \sigma T^4$, where $\sigma$ is the Stefan–Boltzmann constant. Considering the best-fit \texttt{bbody} parameters and their uncertainties from the three models, the resulting apparent radii are $\sim5-9~r_{\rm g}$ and $\sim4-6~r_{\rm g}$ for Obs1 and Obs2, respectively. The inferred emission radii are consistent within the uncertainties between the two observations. Notably, the upper limit of the radius for Obs1 is comparable to the characteristic size of the NS surface, which may suggest that the seed photons originate from a region comparable to the NS surface. On the other hand, the smaller radius inferred in Obs2 is consistent with that reported by \cite{Ursini2023A&A...676A..20U}, suggesting a more compact emitting region. We emphasize that these values represent equivalent blackbody emission radii and do not necessarily correspond to the geometry of the seed-photon emitting region. We also note that we have not accounted for any dilution of the blackbody emission when estimating the BL radius, as the appropriate value of the color-correction factor for persistent BL emission remains uncertain.
 
 The detection of a prominent Fe $K_\alpha$ line together with a reflection hump in the 2019 observation (Obs1) provides the second clear detection of disk reflection in GX~9+9. The first such clear detection was reported by \citet{Ursini2023A&A...676A..20U} using the 2022 NuSTAR observation (our Obs2). Based on the polarimetric analysis, the authors suggested that the fluorescent iron emission line originates from the illumination of the ionized disk by the BL. This interpretation is also consistent with our spectral results, where illumination of the moderately ionized accretion disk by the thermal Comptonization component originating from either the BL or a warm corona produces a narrow to moderately broad iron line. Furthermore, \citet{Iaria2020A&A...635A.209I} demonstrated the need for a reflection component to describe the broadband spectrum of GX~9+9, despite the lack of prominent reflection features. They suggested that the absence of the clear iron line could be associated with the low statistics in the Fe-$K$ band. Given the limited data quality, we find the Fe $K_\alpha$ line profiles in the two observations to be comparable. In contrast, the reflection hump is moderately weaker in Obs1 than in Obs2, which may be related to the larger coronal optical depth and stronger continuum emission in Obs1, potentially diluting the relative contribution of the reflected emission.
 
 The disk inclination angle is poorly constrained by the reflection modeling; nevertheless, the inferred range (see Table~\ref{tab:relxillns}) is in close agreement with the value of $\sim28\degr$ reported by \citet{Ursini2023A&A...676A..20U} and the $\sim39\degr–60\degr$ range reported by \cite{Iaria2020A&A...635A.209I}. It is also worth noting that \citet{Schaefer1990ApJ...354..720S} and \citet{Savolainen2009MNRAS.393..569S} placed an upper limit of $70\degr$ on the disk inclination angle of GX~9+9. The weak reflection features, limited data quality, and degeneracy between inclination and inner disk radius prevent robust constraints on these parameters.
 
 Despite the soft spectral state, both reflection and disk-continuum modeling favor a truncated accretion disk in GX~9+9, with the inner edge located away from the NS surface (see Table~\ref{tab:relxillns} and Section~\ref{sec:analysis}). While the reflection modeling could not constrain the disk radius robustly for Obs1, plausibly due to the weak nature of the reflection features, continuum modeling suggests a relatively smaller radius in Obs2. However, large uncertainties in $R_{\rm in}$ for Obs2 due to limited data quality prevent us from constraining the inner disk radius robustly. Furthermore, the slight difference between the radii estimated from these two methods is simply because \texttt{diskbb} is a non-relativistic disk model and neglects the zero-torque boundary condition at the inner edge of the accretion disk \citep{Mitsuda1984PASJ...36..741M, Shakura1973A&A....24..337S}. Additionally, the inferred reflection fraction, expressed as the solid angle subtended by the reflector ($\Omega/2\pi$), is only $\sim 0.1$ in the 0.7--40 keV band, which further supports a truncated-disk geometry. These results are consistent with the inner disk radius of $5\pm3~r_{\rm ISCO}$, corresponding to $\sim 11$--43~$r_{\rm g}$ for an NS spin of 0.2, reported by \citet{Ursini2023A&A...676A..20U} for the 2022 observation. Furthermore, \citet{Iaria2020A&A...635A.209I} also reported an inner radius of $<15~r_{\rm g}$ for the 2000-2001 observations.

 The accretion disk might be truncated in the soft state of NS-LMXBs if the innermost part of the disk is replaced by an optically thick warm flow or BL. Such a region can also serve as the Comptonizing medium, consistent with recent polarization studies of GX~9+9 \citep{Chatterjee2023MNRAS.521L..74C, Ursini2023A&A...676A..20U}. The relatively large optical depth of the Comptonizing medium inferred from our spectral analysis further supports a Comptonizing region associated with the BL. To investigate whether the disk truncation can be associated with the presence of the BL, we estimated the BL radius, $R_{\rm BL}$, using the relation given by \cite{Popham2001ApJ...547..355P},
 \begin{equation}\label{eq:R-BL}
     \log (R_{\rm BL} - R_{\rm NS}) \simeq 5.02+0.245 \left| \log \left( \frac{\dot{M}}{10^{-9.85}M_\odot \rm yr^{-1}}\right)\right|^{2.19} {\rm cm},
 \end{equation}
 where $\dot{M}$ is the mass accretion rate on the NS surface. For this, we first estimated the mass accretion rate per unit area, $\dot{m}$, on the NS surface using equation 2 of \citep{Galloway2008ApJS..179..360G}. We adopted the persistent flux in the 0.7--40 keV band from model M3 and the bolometric correction factor, $c_{\rm bol}$=1.38, for non-pulsing sources \citep{Galloway2008ApJS..179..360G}. The term $1+z = (1-2GM/R_{\rm NS}c^2)^{-1/2}$, where $z$ is the surface redshift, is 1.31 for a canonical NS with mass $M = 1.4~M_\odot$ and radius $R_{\rm NS}=10$ km. $\dot{M}$ was then calculated as $\dot{M}=\dot{m} \times 4 \pi R^2_{\rm NS}$ g s$^{-1}$. The inferred values of $R_{\rm BL}$ are $\sim17.3$ and $\sim16.6$ km, corresponding to $\sim8.4~r_{\rm g}$ and $\sim8~r_{\rm g}$, for an NS mass of $1.4~M_\odot$ in Obs1 and Obs2, respectively, indicating a comparable radial extent of the BL in both observations. However, these values of $R_{\rm BL}$ appear to be smaller than the inner disk radii inferred for both observations, suggesting that the BL alone may not fully account for the disk truncation. Notably, equation~\ref{eq:R-BL} does not account for the NS spin and viscous effects within the BL, which may affect the accurate estimation of $R_{\rm BL}$ \citep{Mondal2021MNRAS.504.1331M}.

  In NS systems, the stellar magnetic field may also drive disk truncation where the outward magnetic pressure balances the inward ram pressure in the inner disk. Using the measured inner disk radius, $R_{\rm in}$, from reflection modeling, we can place an upper limit on the NS magnetic field strength. For this, we first estimated the NS magnetic dipole moment, $\mu$, following the relation \citep{Cackett2009ApJ...694L..21C},
 \begin{equation}\label{eq:mag}
 \begin{aligned}
    \mu = 3.5\times10^{23}~ {k_{\rm A}}^{-7/4}~ X^{7/4} \left(\frac{M}{1.4~M_\odot}\right)^2 \\ \left(\frac{f_{\rm ang}}{\eta} \frac{F_{\rm bol}}{10^{-9} ~{\rm erg~ cm^{-2}~ s^{-1}}}\right)^{1/2} \frac{D}{3.5 ~\rm kpc} ~\rm G~ cm^3,
    \end{aligned}
 \end{equation}
 where $F_{\rm bol}$ is the bolometric flux, and $M$ and $D$ are the NS mass and distance, respectively. We assumed $k_{A}$, the coefficient that accounts for the conversion from spherical to disk accretion, and $f_{\rm ang}$, the anisotropy correction factor, to be unity. The accretion efficiency, $\eta$, was taken to be 0.1 \citep{Cackett2009ApJ...694L..21C}. The parameter $X$ is related to $R_{\rm in}$ through $R_{\rm in} = X ~GM ~c^{-2}$. We considered the upper limit of $R_{\rm in}$ to be $\lesssim123~r_{\rm g}$ for Obs2, obtained from model M3, which provides proper constraints on the inner disk radius (Table~\ref{tab:relxillns}). For the value of $F_{\rm bol}$ for Obs2 from model M3, the upper limit of $\mu$ is found to be $\lesssim3\times10^{28}$ G cm$^3$. The magnetic field strength, $B$, was then derived from the relation $\mu = B R_{\rm NS}^3$. For $R_{\rm NS}=10$ km, we obtain the upper limit of $B$ to be $\lesssim3 \times 10^ {10} $ G. This is consistent with the upper limit of the magnetic field strength typically observed in weakly magnetized NS-LMXBs \citep[see, e.g.,][]{DiSalvo2003A&A...397..723D, DiSalvo2023arXiv231112516D}.
  
 \citet{Ursini2023A&A...676A..20U} proposed a possible scenario in which the apparent disk truncation could be associated with an additional reflection component originating from the highly ionized innermost disk extending close to the ISCO. Owing to the strong ionization, this reflection component is expected to be featureless \citep{Garcia2022ApJ...926...13G} and can therefore mimic the primary continuum. The authors found that an additional highly ionized \texttt{relxillNs} component, with $\log \xi \approx 4$, could contribute up to $\lesssim 30\%$ of the 2--8 keV flux. However, the limited data quality did not allow them to place strong constraints on this additional component. We also tested this scenario by incorporating an additional \texttt{relxillNs} component into the model M3, fixing $\log \xi = 4$ and setting the inner disk radius to the ISCO. The addition of this component did not result in any significant improvement in the fit statistics or changes in the best-fit parameters. Disentangling a highly ionized reflection component, if present, from the primary continuum will require higher-quality data.
 
 The PDSs and CPDSs from both observations show no sharp features and are dominated by VLFN, likely arising from fluctuations in the mass accretion rate. Such behavior is consistent with that typically observed in bright atoll sources \citep{Lommen2005A&A...435.1005L, vanderKlis2006csxs.book...39V}. The absence of QPO-like features in GX~9+9 has also been reported previously \citep{Wijnands1998IAUS..188..370W}. The observed low-energy variability of $\sim2\%$ is consistent with a relatively stable disk, and high-energy variability of $\sim22\%$ indicates a highly variable Comptonizing component. Such energy-dependent behavior is consistent with that observed in the soft state of NS-LMXBs \citep{vanderKlis2005ASIB..210..283V} and the high/soft state of BHXRBs \citep{Gierlinski2005MNRAS.363.1349G}, where the thermal disk remains relatively stable while the Comptonized component drives most of the fast variability.

 \section{conclusion}{\label{sec:conclusion}}
 Our main results from this study are as follows.

 The broadband spectrum of GX~9+9 in the soft state can be described within the model framework consisting of MCD emission from the accretion disk, thermal Comptonization of seed photons originating from the NS in an optically thick warm flow or BL, and reflection of hard X-rays from the accretion disk. The illuminating spectrum in the form of thermal Comptonization of blackbody seed photons provides an alternative scenario to that of the bulk-motion Comptonization adopted by \cite{Ursini2023A&A...676A..20U}. Moreover, the polarization studies of this source also suggest that the Comptonized emission originates in the BL \citep{Chatterjee2023MNRAS.521L..74C, Ursini2023A&A...676A..20U}.

 While the source remained in the soft state and exhibited a similar spectral shape in both observations, the 2022 observation showed a higher thermal contribution along with a steeper photon index and higher disk seed-photon temperatures than the 2019 observation. The seed-photon radius could be larger, comparable to the size of the NS, in Obs1 than a more compact one, almost half the size of the NS, in Obs2. This is consistent with the relatively higher coronal contribution in Obs1.

 The detection of a prominent Fe $K_\alpha$ line and reflection hump in the 2019 observation provides the second clear evidence of disk reflection in GX~9+9. The iron line is produced by illumination of the moderately ionized disk by thermal Comptonization from the BL or warm corona, in agreement with the polarimetric interpretation of \citet{Ursini2023A&A...676A..20U}. The line strength appears to be similar in both observations within the data quality, while the weaker reflection hump in Obs1 likely reflects dilution by the stronger primary continuum.

 Both reflection and disk-continuum modeling favor the accretion disk to be truncated away from the NS surface, with the latter indicating a relatively smaller disk truncation in Obs2. Our study suggests that the disk truncation in the soft state could be associated with the presence of the BL. On the other hand, if the disk truncation is governed by the magnetic field of the NS, the upper limit on the strength of the magnetic field is found to be $\lesssim3\times10^{10}$ G.

 The observed variability is consistent with a relatively stable disk and a highly variable Comptonizing component, consistent with typical soft-state behavior.

%% Please use the acknowledgment and contribution environments. This will 
%% be anonomyized when the "anonymous" style option is used. 
\begin{acknowledgments}
We thank the reviewer for the constructive comments, which have helped improve the quality of this paper. SC and HKC are supported by the National Science and Technology Council (NSTC) of the Republic of China (Taiwan) under grants NSTC 113--2112--M--007--020 and 114--2112--M--007--042, and by Taiwan Space Agency (TASA) under grants NSPO--P--109221 and TASA--P--1140072. SB acknowledges funding from PRIN MUR 2022 SEAWIND 2022Y2T94C, supported by European Union - Next Generation EU, Mission 4 Component 1 CUP C53D23001330006 and INAF Large Grant 2023 BLOSSOM O.F. 1.05.23.01.13. AAZ acknowledges support from the Polish National Science Center through grant 2023/48/Q/ST9/00138. VKA thanks the GH, SAG; DD, PDMSA; and Director, URSC for encouragement and
continuous support to carry out this research.
\end{acknowledgments}

\bibliography{reference}{}
\bibliographystyle{aasjournalv7}



\end{document}